# Helium Migration through Photomultiplier Tubes – The Probable Cause of the DAMA Seasonal Variation Effect[§]

Daniel Ferenc[1,3,**], Dan Ferenc Šegedin[2,3], Ivan Ferenc Šegedin[3], Marija Šegedin Ferenc[3]

(1) Department of Physics, University of California Davis, Davis CA, USA
(2) Department of Physics, University of California Berkeley, Berkeley, CA, USA
(3) PhotonLab, Inc., Davis, CA, USA

ABSTRACT

The interpretation of the DAMA seasonal variation pattern as a Dark Matter (DM) effect rests on the assumption that all sources of variable background have been excluded from the measurement. We have identified an overlooked background that mimics the DM signature—a large, patently existing effect that has nothing to do with detection within the scintillators. This process takes place exclusively within the individual photomultiplier tubes (PMTs), with two familiar actors (thermionic electrons and helium atoms), two familiar processes (helium penetration of glass and ion afterpulsing), and a simple storyline: (a) After defeating the insulation of the DAMA detector, helium atoms from the variable local environment penetrate the PMTs' vacua; (b) Thermionic electrons from the PMT photocathode ionize helium atoms; (c) Secondary electrons from each helium ion's impact in the photocathode form an afterpulse, and thus effectively boost and extend the waveform. We demonstrate that the accidental coincidences of such inflated dark-noise waveforms, originating from any pair of PMTs, mimic scintillation events at a rate of ~1 cpd/kg/keV, which is consistent with DAMA's result. The reported seasonal variation of ~1% thus implicates a variation of the helium concentration in the local environment that can cause a ~1% modulation of the helium pressure within the PMTs. We predict that the DAMA detector with a modified readout logic – blinded for the DM-like scintillation events – will measure the same, helium-driven seasonal variation pattern as the original detector.

---

[§] This is **Version-2** of the article that includes additional information in order to emphasize the significant extent to which helium pressure can vary in typical mining and geological settings. We also further clarify the "equilibration" process of the helium pressure within the PMTs. We propose measurements to monitor key parameters associated with *TheHeap* mechanism.
[**] Corresponding author, ferenc@physics.ucdavis.edu or photonlab@gmail.com

## 1. INTRODUCTION

The DAMA Experiment, located at the INFN Gran Sasso National Laboratory (LNGS), has reported highly significant seasonal variation in the event rate detected by 25 ultrapure NaI(Tl) scintillators, each of them viewed by a pair of photomultiplier tubes (PMTs) [1]. This result has been interpreted as a model-independent signature of the interactions of hypothetical Dark Matter (DM) particles within the scintillators. The observed modulation is consistent with the predictions based on the motion of the Earth relative to the galactic halo.

Partly motivated by the unique promise of this model-independent DM detection method, we have recently engaged in a conceptual design of generic next-generation DM detectors (in cryogenic liquids, scintillators, and even in ice [2]), which would take full advantage of the advanced ABALONE$^{TM}$ Photosensor Technology [3]. This modern technology was specifically invented and developed for cost-effective mass production, robustness and high performance (intrinsically high gain in the upper $10^8$ range, single-photon sensitivity and resolution, UV-sensitivity as well as unique radio-purity thanks to fused-silica components - at no additional cost to the assembly process, and helium afterpulsing rate about two orders of magnitude lower than in PMTs, restricted to a very narrow time interval) [4,5]. These features provide significant advantages over the 80-year old PMT technology.

We have approached this design task from the perspective of scientists who design and operate various ultrahigh vacuum (UHV) systems and instrumentation for research, as well as for production and evaluation of photosensors. One of our primary concerns has been helium gas, particularly the possibility of its uncontrolled release in a wide perimeter around our experimental and manufacturing facilities. In that sense, deep underground laboratories are of the highest concern due to natural fluctuations in the concentration of various gases, including radioactive radon, and seemingly harmless helium. In the vast majority of underground research laboratories and mines, the concentration of radon has been closely monitored. Helium, on the other hand, has largely been ignored as it is neither radioactive nor chemically active. Standard mining and geological studies, however, have tracked helium variation and used it as a key research tool [6,7]. Very high helium concentration (e.g. 20 times higher than the average surface concentration [6]), as well as positional and seasonal variations, should be considered as a given in underground laboratories. Moreover, in contrast to radon or any other gas, helium cannot be efficiently ventilated. It is extremely intrusive and penetrates tight enclosures with ease. In UHV systems, the apparent absence of helium is merely due to the dynamic equilibrium between constant pumping and helium penetration, which occurs both through the system structure, and backward through the vacuum pumps.

Helium is thus the only element from the local variable environment that can penetrate the best detector shields like those used by DAMA. Once inside the detector, helium slowly penetrates the glass enclosure of any PMT at a rate proportional to the pressure difference between the inside and the outside, as well as PMT characteristics such as constructional materials, size, shape, and the design of the feedthroughs. Then, once inside the PMTs, helium atoms start to play an active role – as ions. The constant stream of energized thermionic electrons from the photocathode must ionize residual gas constituents, including helium (mechanism discussed in detail in section 2). The helium-ion afterpulsing rate must therefore reflect the concentration of helium in the local environment. Therefore, already early on during our studies, we realized that the unprecedentedly low afterpulsing rate of the ABALONE[TM] Photosensors (furthermore strictly confined to a very narrow time-of-flight interval) should play a key role in the next generation of DM experiments. Nevertheless, after a deeper investigation, we were surprised by the magnitude of the effect caused by helium afterpulsing.

We found that – specifically for PMTs – the boosted amplitude of the helium-ion afterpulsing events, in conjunction with their long effective duration that is coincidentally equivalent to the decay time of the NaI(Tl) scintillator, makes these dark-noise events virtually indistinguishable from the waveforms of events that would originate in the scintillators. These events thus mimic the hypothetical Dark Matter events. The T̲hermionic e̲lectron-induced H̲e-ion a̲fterp̲ulsing *(TheHeap)* mechanism is therefore the source of the predominant and uncontrollable variable background. Consequently, a deep underground experiment designed to operate at room temperature and with only two PMTs per scintillator, could reach at best a marginal sensitivity to rare events like DM.

The subject of this article is an important corollary—we conclude that the DAMA detector, which uses strongly afterpulsing PMTs, only two PMTs per scintillator, and operates almost at a room temperature (+18°C), must have been dominated by the *TheHeap* mechanism.

As discussed in more detail below, the design of the DAMA detector has three major problems: (i) the assumption that the pulses of all dark-noise events are necessarily short and weak; (ii) no recognition of the ability of helium from the local environment to penetrate the PMTs' vacua, and (iii) no recognition of the crucial role played jointly by thermionic electrons and helium atoms within the PMTs in the creation of dark noise that mimics the DM signal.

The presented analysis of the *TheHeAP* mechanism in the DAMA experiment is based on numerical simulations of ion optics in a representative PMT model. First, we will show that the timescales and the amplitudes of the complex waveforms that comprise the initial thermionic electron pulses and the ion afterpulses can pass DAMA's event-selection criteria. We will then show that the rates of such inflated DM-like dark-noise events are consistent with the observed phenomenon.

Note that we do not comment on the other attempts to give alternative explanations to the DAMA effect, because virtually all of them assume a detection of real signals in the scintillators. We assert that scintillators play no role in the observed effect.

## 2. *TheHeap* MECHANISM

Once inside a PMT vacuum, each helium atom presents a potential ionization target for the energized thermionic electrons that emerge from the photocathode at a constant rate. Each helium ion generated at any point on the electron trajectories (Fig.1) accelerates toward the photocathode (Fig. 2). Several secondary electrons are then ejected upon the impact of the helium ion in the photocathode. Using numerical simulations, we have determined that the time-of-flight of $He^+$ and $He^{++}$ ions originating from various points within the PMT can range between 200 and 1000 ns. Crucially, ions of other light atoms and molecules take significantly more time to reach the photocathode (Table 1).

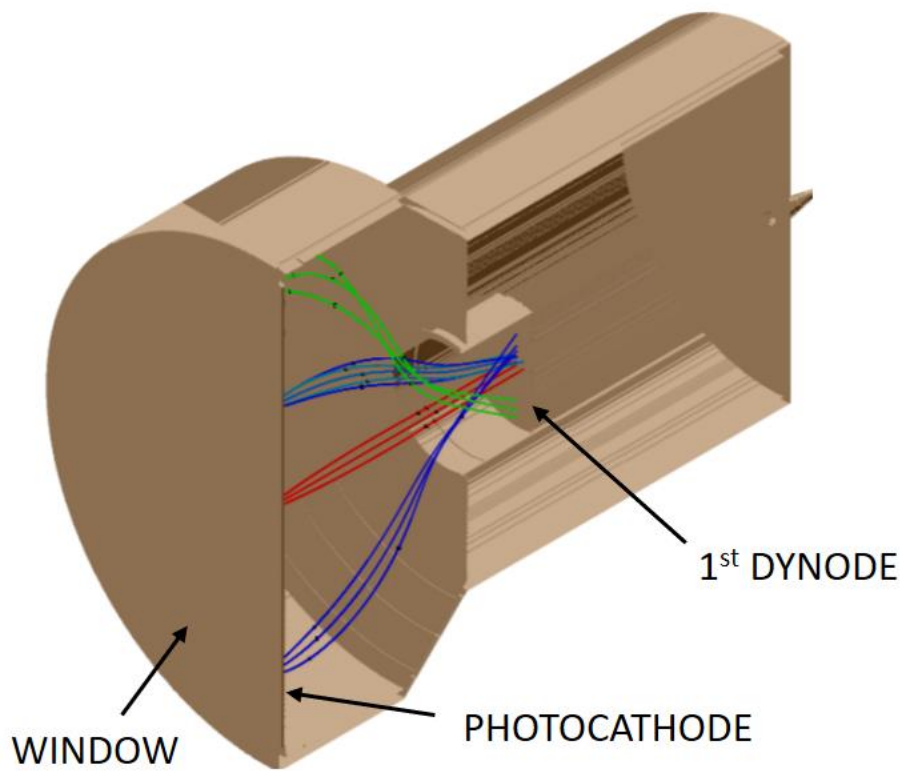

Fig. 1 Electron optics of a PMT model representing the 3-inch PMTs used in the DAMA detector [8], implemented in the SIMION ion optics simulation program. Trajectories represent either photoelectrons or thermionic electrons originating at various points on the photocathode surface, focused into the first dynode area. The potential between the photocathode and the first dynode was assumed to be 350 V [8]. Note that the rest of the dynode chain was of no interest to our study.

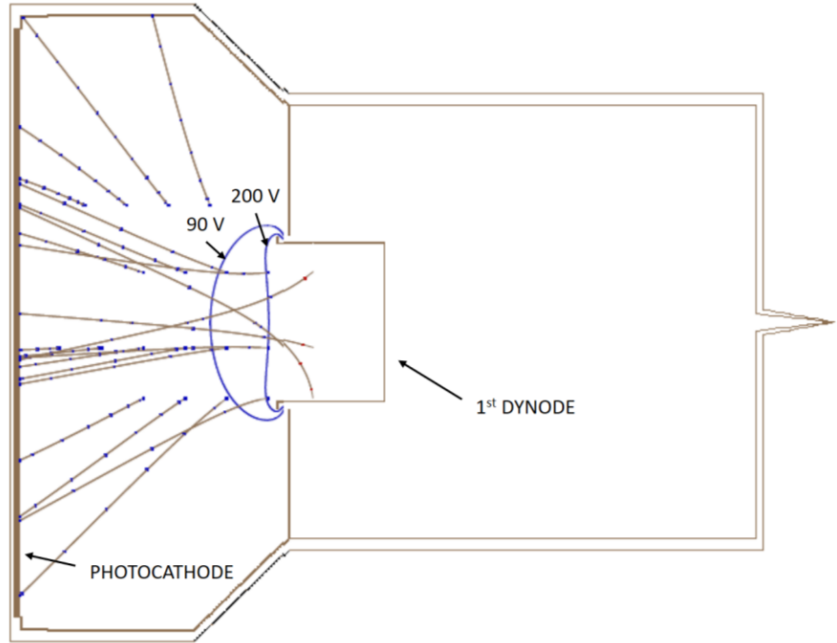

Fig. 2 Simulated trajectories of He$^+$ ions on the way from their origins along the photoelectron trajectories to the photocathode (i.e. traveling from right to left). Marks along the trajectories correspond to intervals of 100 ns. The ionization probability peaks for ions generated between the two indicated equipotential lines [9].

Table 1. Time-of-flight (TOF) of the afterulses of various light ions, assuming a potential of 350 eV between the photocathode and the first dynode (Fig. 2). Note that the TOF range of helium ions is comparable to the decay time of NaI(Tl) scintillator (240 ns), which is why *TheHeap* event coincidences can mimic scintillation events.

| ION | TOF range [ns] | |
|---|---|---|
| He$^{++}$ | 219 | 683 |
| He$^+$ | 310 | 967 |
| C$^+$ | 537 | 1674 |
| N$_2^+$, C-O$^+$ | 821 | 2560 |

In the ion optics simulations (using the SIMION program), we implemented the dimensions of the Hamamatsu R6233 PMTs used by DAMA (Figure 2 in Ref. [8]), and fixed the high voltage between the photocathode and the first dynode to 350 V. Note that the high voltages applied to the 50 PMTs in DAMA range from 0.9 to 1.3 kV [8], in order to equilibrate the individual responses (nevertheless, the range of sensitivity remained between 6 and 10 electrons per keV of energy deposited in the 25 scintillation detectors). According to the schematic of the voltage divider (Figure 6 in Ref. [8]), this range translates to the range of 300-433 V for the voltages between the photocathode and the first dynode. Our choice of 350 V is approximately in the middle of that wide interval.

A closer inspection of the time marks along the ion trajectories in Fig. 2 reveals that ions that travel longer distances arrive to the photocathode faster; this is because the potential difference with respect to the photocathode is higher at stating points farther from the photocathode. Since the ionization efficiency strongly peaks for electron energies of about 90-200 eV [9], the vast majority of helium ions should have relatively high energies and short times of flight. In addition, more energized helium ions lead to higher afterpulsing amplitudes and are thus more likely to be considered DM candidates. The first half of the TOF range between 200 ns and 1000 ns should therefore be significantly more populated than the second. In the following calculations, we will focus on the first half of that interval.

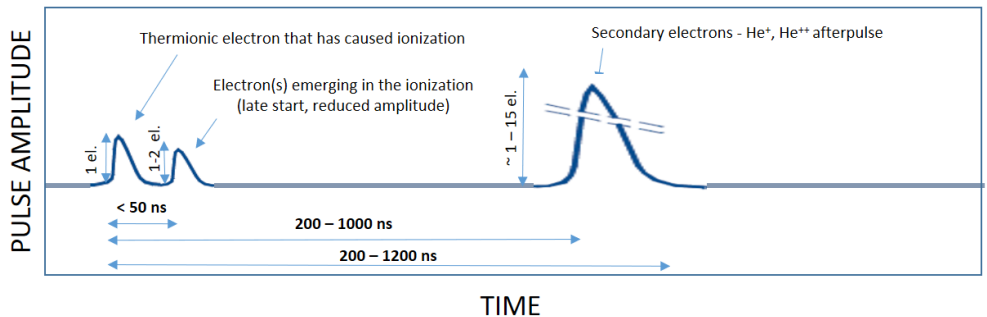

Fig. 3 Schematic view of a typical *TheHeap* event. An energized Thermionic electron (*The*) from the photocathode ionizes a helium atom, and continues towards the first dynode, in which it initiates the first pulse. Shortly after that, one or both electrons from the ionization of the helium atom lead to the second pulse. That pulse is in general weaker, because these electrons start in a weaker electric field, somewhere between the photocathode and the dynode. In reality, these two pulses mostly fuse into a single pulse, which is thus more likely to pass the hardware trigger than a single-electron pulse. A positive helium ion generated in the ionization process accelerates towards the photocathode, and releases secondary electrons upon its impact. Those electrons are accelerated onto the first dynode, and their detection leads to an afterpulse in the waveform.

The primary reason for the high afterpulsing rate in the PMTs — the key problem discussed in this article — is precisely in the fact that the peak of the electron-impact ionization cross section overlaps with the actual electron energies in the PMTs. This is not by accident, as the high voltage between the photocathode and the first dynode in virtually all PMTs is purposely chosen to maximize the ionization effect, not in the residual gas, but rather to optimize the rate of secondary electron emission from the first dynode. The ionization cross section of helium falls from the peak value by a factor of 20 at 10 keV, and by about two orders of magnitude at 20 keV. (The low ionization probability at high voltage is one of the reasons why the ABALONE[TM] Photosensors (which operate at ~20 keV) have a significantly lower afterpulsing rate than any PMT [4,5].)

The detection of secondary electrons that emerge from the photocathode upon the ion impact leads to an afterpulse, i.e. a pulse in the waveform that follows (1) the pulse of the original thermionic electron, and (2) the pulse(s) of one or two electrons released from the helium atom in the ionization process. The delay of the afterpulse corresponds to the time-of-flight of the ion. Note that the starting signal itself can also be extended, because the electron(s) released in the ionization process are delayed by up to about 50 ns. The afterpulses can also be dispersed in time due to the spread in the initial momenta and angles of the secondary electrons. The resulting waveform therefore starts with a 10-50 ns-wide, 1-3 electron-strong peak, and ends with a ~1-15 electron-strong configuration that arrives 200-1000 ns later and lasts up to about 1200 ns (Fig. 3). As discussed above, the majority of the accepted ionization events are in the 200-600 ns range.

The *TheHeap* mechanism thus effectively both stretches (from ~10 ns to between 200-1200 ns), and amplifies (~2-15 times) the initial thermionic single-electron signals. Although generated exclusively within the PMTs, these inflated dark-noise signals can be mistaken for real scintillation signals. Their accidental pairwise coincidences (both generated and detected by pairs of PMTs) can pack ~2-6 electrons in the initial group, and ~2-30 electrons in both afterpulses (i.e. all together ~4-36 electrons, over a period of ~200-1200 ns). As demonstrated below, a significant fraction of such dark-noise events can pass DAMA's criteria as valid DM event candidates.

### 3. DAMA'S EVENT SELECTION METHOD

The two PMTs connected to each scintillator work in coincidence. The hardware threshold used by the experiment is set to a "single photoelectron level," while the software energy threshold is set at 1 keV [1]. The hardware selection of the coincident events takes place within a 100 ns time window, as illustrated with a coincidence box shown in the upper two sections of Figs. 4-8. The lower sections illustrate the next step in the selection process, discussed below in some detail.

In the illustrations in Fig. 4 and Fig. 7, the trigger comes from the overlap of the initial electron groups in both PMTs, while in Fig. 5 and Fig. 8, a solitary thermionic electron from one PMT overlaps with the initial electron group from the other. In Fig. 6, the initial electron group from one PMT overlaps with an afterpulse from the other. The example shown in Fig. 8 illustrates a $CO^+$ ion afterpulse, whose time-of-flight must be significantly longer than that of helium ions (Table 1). Note that in reality the afterpulses shown in Figs. 4-8 may be structured, while the initial electron groups are likely to fuse into single peaks, which makes the passage of the hardware trigger more likely.

Once the coincident event passes the hardware selection criterion, the purpose of the subsequent steps in DAMA's analysis has been to select potential DM-induced recoil events in the scintillators, and reject accidental coincidences of dark-noise events. The signal amplitudes of the low-energy events close to the detection threshold (of interest for DM detection) are not high enough for a precise analysis of exponential decay curves and their comparison to the decay time of the NaI(Tl) scintillator. Instead, the signals from both PMTs are first combined, and then subjected to an empirical pass-fail cut, based on two parameters that characterize the evolution of the combined waveform. This event selection process is outlined in Ref. [8];

> "… the NaI(Tl) scintillation pulses (time-decay ~240 ns) [are] well distinguishable from the noise ones (time decay ~tens ns). (…) The different time characteristics of these signals can be investigated building several variables from the information recorded by the Waveform Analyzer over the 2048 ns time window. The variables follow different distributions for scintillation and noise pulses. In particular, we consider the fractions of the pulse areas evaluated over different time intervals, such as:
>
> X1 = Area(from 100 to 600 ns) / Area(from 0 to 600 ns)
>
> X2 = Area(from 0 to 50 ns) / Area(from 0 to 600 ns).
>
> The first variable is distributed around 0 for noise events and around 0.7 for scintillation pulses; the second variable indeed is distributed around 1 for noise and around 0.25 for scintillation events."

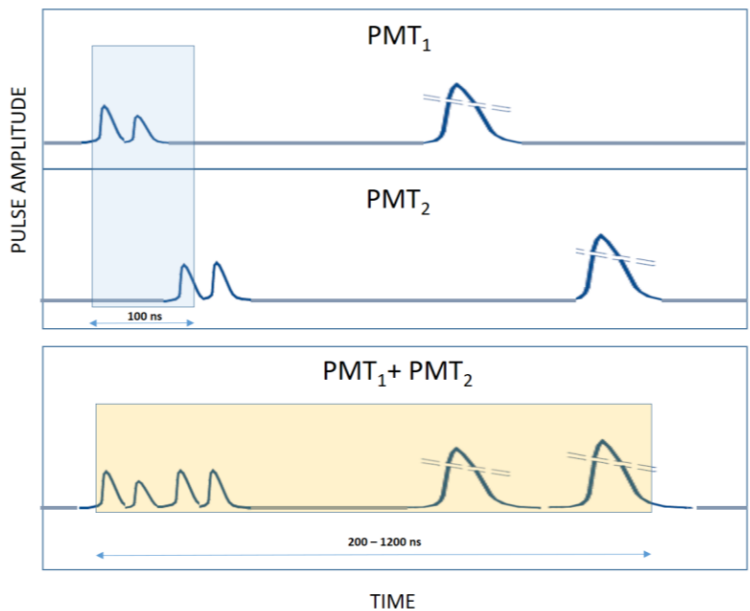

Fig. 4 An illustration of an accidental coincidence of two independent *TheHeap* events in PMT$_1$ and PMT$_2$. Both PMTs oversee the same NaI(Tl) scintillator, but the scintillator plays absolutely no role in the *TheHeap* mechanism. In this particular case, the overlap of the two initial pulse groups within 100 ns would trigger the coincidence in DAMA's analysis.

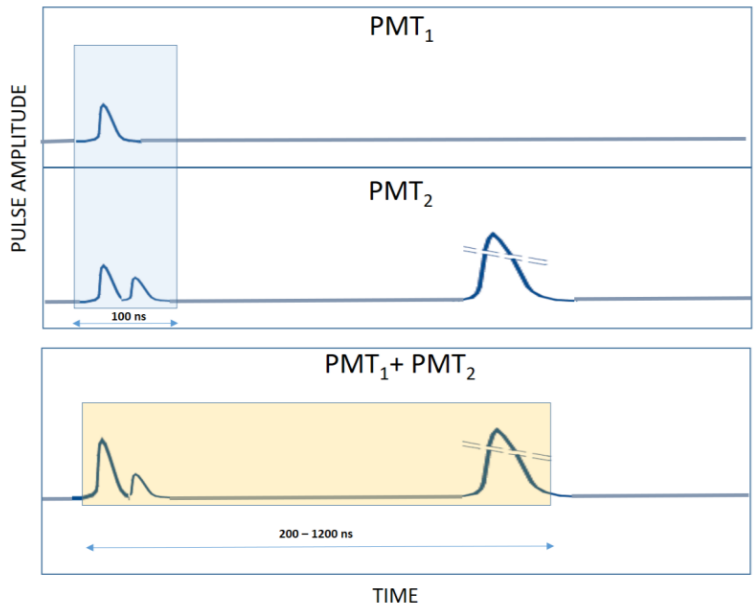

Fig. 5 Coincidence of a solitary thermionic electron event (*The*) in $PMT_1$, and an independent *TheHeap* event in $PMT_2$.

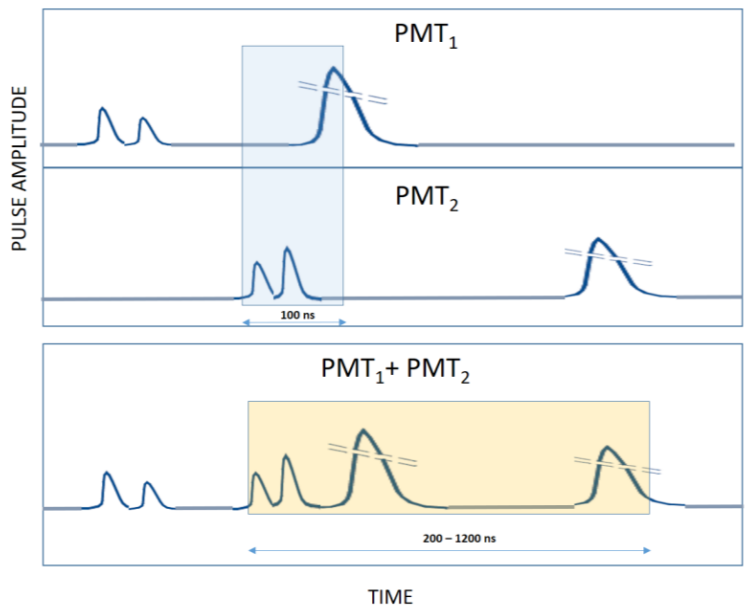

Fig. 6 In this example, the coincidence trigger is caused by the initial pulse group (*The*) in $PMT_2$, and the afterpulse from $PMT_1$.

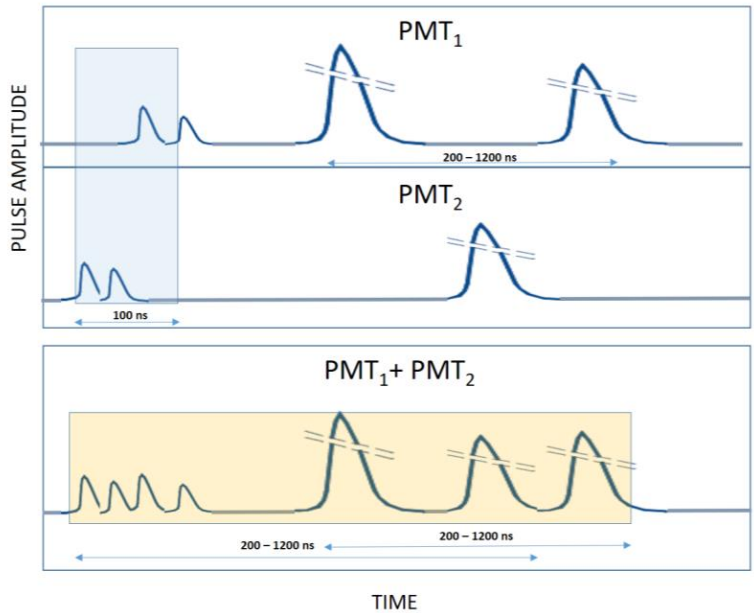

Fig. 7 In $PMT_1$ a secondary electron within the afterpulse initiates another afterpulse. Such second-order events are somewhat less likely, but they can have high amplitudes.

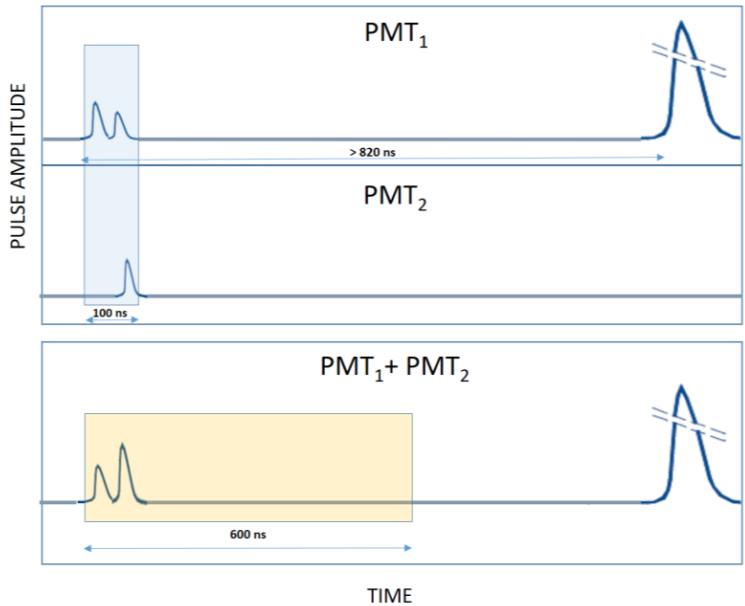

Fig. 8 Coincidence of a Thermionic electron-induced $CO^+$-ion afterpulsing event (*TheCOap*) in $PMT_1$, with a solitary thermionic electron event (*The*) in $PMT_2$. Since the delays of the afterpulses of atoms and molecules heavier than helium are significantly longer than 600 ns (see Table 1), they cannot pass DAMA's X1-X2 event selection criteria as scintillation events and DM candidates.

To illustrate how the X1-X2 event selection method works with the *TheHeap* events, we analyze the five examples shown in Figs. 4-8, assuming that the total timespan of each event (the length of the box in the lower section of the figures) is 600 ns, for the reasons explained in the previous section. Table 2 shows the results, including the "energies" that would be associated with the number of electrons in DAMA's analysis. Since in reality these electrons are not photoelectrons, and the events have nothing to do with the scintillator, the associated "energy" has no physical meaning. However, we have to define it for the comparison with the normalized event rates reported by DAMA.

Table 2. Properties of the illustrative examples shown in Figs. 4-8.

| Figure | Event Type | Number of electrons | Associated "Energy" (scaling factor = 6 electrons/keV) | X1 | X2 |
|---|---|---|---|---|---|
| Fig. 4 | *TheHeap-TheHeap* | 12.5 | 2.1 keV | 0.8 | 0.1 |
| Fig. 5 | *The-TheHeap* | 7.5 | 1.25 keV | 0.7 | 0.3 |
| Fig. 6 | *TheHeap-TheHeap* | 11 | 1.8 keV | 0.8 | 0.2 |
| Fig. 7 | *TheHeap-TheHeap+Heap* | 18.5 | 3.1 keV | 0.9 | 0.1 |
| Fig. 8 | *TheCOap-The* | 2.5 | 0.2 keV | 0.0 | 0.6 |

The first four examples satisfy DAMA's expectations for scintillation events (X1~0.7; X2~0.25), and all of them are inconsistent with DAMA's definition of a "noise event" (X1~0; X2~1), although they are in fact dark-noise events. Moreover, all of these examples pass the 1 keV "energy" threshold, and thus qualify as DM event candidates.

In contrast to the four cases that involve helium ion afterpulses (Figs. 4–7), the illustration of a $CO^+$ ion afterpulse (Fig. 8) is inconsistent with a scintillation event, and consistent with noise, because its effective width exceeds the 600 ns time interval (Table 1). It is essential that only the helium afterpulsing waveforms can mimic scintillation events, because their typical widths are coincidentally equivalent to the decay time of NaI(Tl) scintillators. Unfortunately, helium is also the only element from the variable environment that penetrates the PMTs. The rate of dark-noise events that mimic scintillation events must therefore be correlated with the variable pressure of helium in the local environment.

## 4. THE EVENT RATE

The purpose of the following analysis is to investigate whether various types of *TheHeap* coincidences can occur frequently enough to explain the reported event rate.

The canonical value for the rate of Thermionic electron *(The)* emission from bialkali photocathodes at room temperature is 50 Hz/cm$^2$: the 3" PMTs used by DAMA would thus have a total rate of about 2.3 kHz. At a somewhat lower operating temperature (18°C), and using the photocathode material that has been optimized for low noise, the actual noise rate should be lower; we estimate between 500 Hz and 1.5 kHz.

Note that DAMA's estimates for the noise rate range between 25 and 550 Hz for the 50 PMTs used in the experiment (Figure 8 in [8]). These measurements are mutually inconsistent. Moreover, they contradict the physics of thermionic emission in magnitude (as a systematic underestimate) and dispersion. In general, the effective rate in the actual measurements strongly depends on the precise definition of the hardware threshold, specified by DAMA as "at the single photoelectron level" [8], i.e. on the precise position of the acceptance threshold between the pedestal and the single-photon peak, the region usually referred to as the valley. Note that the wider valley region, including the pedestal peak, is populated by electrons that have back-scattered from the first dynode, after converting a fraction of their energy into secondary electrons. However, by reaching the first dynode, these partially detected electrons were equally productive in generating afterpulses as the fully detected electrons. Moreover, as noted above, the initial electron groups in afterpulsing events in general fuse into single peaks of increased amplitude. That allows the afterpulsing events to pass the trigger threshold more likely than the events with solitary thermionic electrons. In the absence of a reliable experimental acceptance rate of thermionic electrons, let alone the afterpulsing events, we will use the lower limit of our estimated range, i.e. 500 Hz, which still coincides with the uppermost reported level. This is a conservative estimate for the lower limit of the actual acceptance rate.

The probability for a random *The* event to appear in a single 100 ns-wide time interval (trigger window) is $r_{The}$ ~ 500 Hz x 100 ns = $5\times10^{-5}$, and the double-hit probability (*The-The* coincidence in a pair of PMTs) within the same 100 ns interval is $r_{The-The}$ ~ 2 x $(r_{The})^2$ = $5\times10^{-9}$.

If one of the two coincident thermionic electrons is followed by an afterpulse (like in *TheHeap* event illustrated in Fig. 3), the afterpulsing probability should also be taken into account. We assume an afterpulsing rate of ~0.05 helium ions per thermionic electron, which is again a conservatively low estimate for a PMT of this size [10]. The rate of *The-TheHeap* events (Fig. 5) is thus $r_{The-TheHeap}$ ~$2.5\times10^{-10}$, still referring to a 100 ns trigger interval. Likewise, the rate of *TheHeap-The*ap coincidences (Fig.4) is $r_{TheHeap-TheHeap}$ ~ $1.25\times10^{-11}$.

Renormalized to a daylong period, the *TheHeap-The*ap rate is 11 cpd (counts per day). Furthermore, the event rates are usually presented in a normalized form, i.e. as cpd divided by the scintillator mass and the width of the "energy" interval. In our case, neither of these two quantities makes physical sense, since *TheHeap* effect takes place only within the PMTs. But at the end of this exercise we will compare our predictions to DAMA's measurement results in the normalized form. The mass of each of the 25 scintillators monitored by a pair of PMTs is 9.70 kg, and we will refer to a 2 keV wide "energy" window (e.g. 1-3 keV), which results in a rate of ~0.6 cpd/kg/keV.

This rate includes events of all amplitudes, i.e. before the cut at the 1 keV "software energy threshold" (i.e. ~6 electrons, including the two initial electron groups). This threshold is low enough for the majority of the *TheHeap-TheH*eap events to pass. Only events in which both afterpulses contain just one electron will almost certainly fail, i.e. an estimated 15-25%, which leaves the event rate at approximately 0.5 cpd/kg/keV.

Among *The-TheH*eap events, on the other hand, only those who's single afterpulse exceeds about ~3-4 electrons will be able to pass the 1 keV "software energy threshold." The distribution of the afterpulsing amplitudes is in general a steeply falling function; we therefore estimate that not more than ~5% of *H*eap events will satisfy this requirement. This brings the rate of *The-TheH*eap events to ~0.5 cpd/kg/keV, just as of *TheHeap-TheHeap* events. The total rate of all *TheHeap* configurations is thus on the order of 1 cpd/kg/keV, in agreement with the published results [1]. In addition, events of the types illustrated in Fig. 6 and Fig. 7 should also contribute to the total event rate. The distribution of the total measured rate as a function of "energy" (shown in Fig. 1 in Ref. [1]) falls from 1.6 cpd/kg/keV at the 1 keV threshold, down to 0.8 cpd/kg/keV at 3 keV, and finally down to 0.3 cpd/kg/keV at 6 keV. The 0.3 cpd/kg/keV level persists to the end of the scale at 10 keV.

The amplitude of seasonal variations reported by the DAMA Collaboration [1] is 0.0103 +/- 0.0008 cpd/kg/keV, i.e. about 1% of the total event rate. However, the variation effect gradually fades out towards the "energy" of ~6 keV. Our explanation for that decline is based on the gradually falling afterpulsing amplitude distribution: because afterpulses should very rarely exceed 15 electrons, *TheHeap-TheHeap* coincidences are limited below ~30 electrons, i.e. below an "energy" of ~5 keV.

The DAMA Collaboration has also reported the rate of coincident events in two scintillators and pointed out that this rate is not seasonally variable. Their interpretation of the absence of variation relies on the assumption that DM particles can interact only in individual scintillators. From our perspective, a double-scintillator coincidence is just a quadruple PMT coincidence. The rate for any quadruple coincidence, e.g. *TheHeap-TheHeap-TheHeap-TheHeap*, is indeed negligible.

As illustrated in Section 3 (Fig. 8 and Table 2), the coincidences that comprise afterpulses of carbon, nitrogen and oxygen (as well as of various molecules, like CO, $CO_2$ and $CH_4$), feature timescales (Table 1) well above 600 ns. Their chances of passing the X1-X2 selection criteria are therefore slim. The assumption that the bulk of the event rate is mainly due to helium afterpulses thus seems plausible. If we assume that the helium pressure within the PMTs has increased by 0.5%, the afterpulsing rate should increase from 0.05 to 0.05025 helium ions per thermionic electron. This leads to an increase in the rate of *TheHeap-TheHeap* coincidences by 1%, a result consistent with the measurements [1].

To illustrate the robustness of our result, we assume large variations in both rates. For the rate of *The* events, we assume 500 Hz < $r_{The}$ < 1500 Hz, which results in coincidence rates of 0.5 < $r_{TheHeap\text{-}TheHeap}$ < 4.5 cpd/kg/keV, and also 0.5 < $r_{The\text{-}TheHeap}$ < 4.5 cpd/kg/keV. Factor of two variations (up and down) in *Heap* (i.e. $0.025 < r_{Heap} < 0.1$ ions/electron) lead to $0.125 < r_{TheHeap\text{-}TheHeap} < 2$ cpd/kg/keV, and $0.25 < r_{The\text{-}TheHeap} < 1$ cpd/kg/keV. These variations lie within the ballpark of the published measurements. Given the large variation in the sensitivity of the individual PMTs and the imprecise definition of the hardware threshold "at a single electron level" [8], a more precise comparison would be very hard.

Let us stress that *TheHeap* events are inevitable, while their event rate can vary to some degree. Quite remarkably, the DAMA detector can – on its own – precisely measure the amplitude of *TheHeap* contribution, after only a minor modification to the readout logic. By merely reconnecting 25 out of the 50 signal cables from the PMTs, one can establish pairwise coincidences of PMTs that are optically coupled to different scintillators, instead of the same scintillators (alternatively, one can introduce delay lines). That way, the DAMA detector will become blind to the DM-like scintillation events, while it will remain equally sensitive to *TheHeap* events. We predict that this explicitly DM-blind version of the DAMA detector will find the same seasonal variation pattern as the original DAMA detector, since in both cases the seasonal effect is the consequence of variations in helium rather than DM. We also predict that the variation amplitude should be measurably higher, because some background events from the scintillators and from the radioactive decays in the PMT glass, which have previously diluted the variation effect, will now be excluded. This will allow DAMA to disentangle these contributions from *TheHeap* events. To further disentangle scintillation noise, and the PMT noise other than *TheHeap*, the PMTs should be completely blinded off the scintillators.

In addition, we recommend comprehensive monitoring of the helium concentration in the local environment, as well as within the detector. We also recommend monitoring the helium concentration within each PMT, both by periodic time-of-flight mass spectroscopy, and by studying the raw event rate systematics as a function of time (constant migration of helium into the PMTs should result in a steady rise of the event rate on top of the seasonal modulation).

## 5. DISCUSSION

Since the *TheHeap* mechanism is driven by two factors – helium ion afterpulses and the thermionic electrons that initiate them – there are two straightforward ways of suppressing this effect.

First, strong suppression of thermionic emission is possible with the reduction of temperature, approximately by factor of 10 for every $15^0$ C (down to a temperature of about $-10^0$ C, below which other electron emission mechanisms dominate in PMTs). Therefore, we expect that the DM-Ice detector, configured similarly to DAMA but placed deep in the South Pole ice, should not find any significant seasonal variation effect, even if some strong (although unlikely) helium fluctuations were to be present at that peculiar site. Indeed, the DM-Ice Experiment [11] has recently reported the absence of seasonal variations, which further corroborates our assertion that the DAMA Experiment has detected seasonal variation of helium rather than DM interactions in its scintillators.

Second, the suppression of the helium afterpulsing rate per thermionic electron can only be achieved with modern photosensors. As discussed above, the ABALONE$^{TM}$ Photosensors [4, 5] provide a very low intrinsic afterpulsing rate, all strictly contained within a very narrow time-of-flight interval (e.g. 120-180 ns for the configuration presented in Ref. [4], which is indeed much shorter than the decay time of NaI(Tl)). We expect at least a four orders of magnitude suppression of the coincident *TheHeap* events, which will grant a room-temperature DAMA-like detector with significantly higher sensitivity.

Operating a detector in a low-temperature environment is clearly an effective way of suppressing the *TheHeap* mechanism. However, it is rather impractical. For a DAMA-like experiment located in a deep underground area, a significant temperature reduction would require highly non-trivial additions of new components and materials. We conclude that this would be useful, but not necessary, given the emerging technology that provides ultralow-afterpulsing, ultralow-radioactivity, and single photon-resolving photosensors.

## 6. SUMMARY

The DAMA Collaboration has stated recently [1] that *"(n)o systematics or side reaction able to mimic the exploited DM signature (…) has been found or suggested by anyone throughout some decades thus far."* We report on an overlooked background mechanism that indeed mimics the DM signature. This mechanism is a patently real, large, first-order effect. The reason it has remained unidentified for such a long time is likely that it has nothing to do with the detection of real events in the scintillators. This process takes place exclusively within the individual PMTs, and presents an ordinary event in the life of every PMT: only two, very familiar actors play an active role (thermionic electrons and helium atoms), only two, very familiar processes are at work (helium penetration of glass, and ion afterpulsing), and the storyline is simple:

i. After defeating the insulation of the DAMA detector, helium atoms from the variable local environment penetrate the PMTs' vacua.
ii. Thermionic electrons from the photocathode ionize helium atoms and other residual gases within the PMTs.
iii. Ions are accelerated towards the photocathode.
iv. Secondary electrons emitted upon each ion's impact in the photocathode form an afterpulse in the waveform, which arrives with a delay after the initial thermionic electron.
v. The delay of the afterpulse is characteristic of the mass and the charge of the ion. Thanks to the low mass, helium afterpulses are delayed much less than those of any other constituents of residual gas in the PMTs, except of hydrogen. Decisively, the range of the helium time-of-flight is coincidentally equivalent to the decay time of the NaI(Tl) scintillator.

The elongated and boosted waveforms of such dark-noise events that contain helium afterpulses precisely mimic scintillation events in DAMA's event selection process. We demonstrate that a large fraction of accidental coincidences of such inflated waveforms from any pair of PMTs must pass DAMA's criteria as valid DM candidates.

With conservative assumptions for the rate of thermionic emission and afterpulsing, we estimated the total rate of events that pass DAMA's selection criteria to be ~1 cpd/kg/keV. Our estimate is consistent with DAMA's published result, within the wide uncertainties due to the variation in the properties of the 50 PMTs. Let us stress that these, previously overlooked events, simply must be present, while the event rate can vary to some degree.

Since helium atoms penetrate PMTs at a rate nearly proportional to the pressure difference, the measured event rate must reflect the concentration of helium in the local environment. The reported seasonal variations of ~1% [1] therefore implicate seasonal variation of the helium pressure in the local environment that was capable of causing a ~1% modulation of helium pressure within the PMTs.

With a trivial modification to the readout logic (merely a reconnection of 25 out of the 50 signal cables from the PMTs), the DAMA experiment can—on its own—measure the amplitude of helium variations. By establishing pairwise coincidences of PMTs that are optically coupled to different scintillators, instead of the same scintillators, the DAMA detector will become blind to the DM-like scintillation events, but remain equally sensitive to the helium-modulated events. We predict that the DM-blind version of the DAMA detector will find the same seasonal variation pattern as the original DAMA detector (the amplitude of variations should be higher, because some background events from the scintillators that have previously diluted the variation effect, will now be excluded). We would expect the same result even if the PMTs were blinded off the scintillators.

Furthermore, we claim that the similarly designed DM-Ice Experiment at the South Pole should find no seasonal variations, because thermionic emission is suppressed at low temperatures (even if some unlikely strong helium fluctuations were to be present at that peculiar site). Indeed, the DM-Ice Experiment [11] has recently reported the absence of seasonal variations.

Our assertion that the DAMA Experiment has actually detected seasonal variations of helium rather than DM interactions in its scintillators is thus further corroborated by the simultaneous presence of seasonal variations in the original DAMA measurements, and their absence in DM-Ice.

## 7. CONCLUSION

While working on the conceptual designs of next-generation Dark Matter (DM) experiments based on the advanced ABALONE[TM] Photosensor technology [3-5], we identified a source of background in the DAMA Experiment (LNGS) that has apparently remained undiscovered [1]. The Thermionic electron-induced He-ion afterpulsing mechanism (*TheHeap*), which is based on patently existent processes that take place within any PMT (and has nothing to do with the scintillators), mimics scintillation signals at a rate comparable to the measured total event rate [1].

We assert that the DAMA Experiment has actually detected seasonal variations of the helium partial pressure in the local environment, rather than DM interactions in scintillators. We describe how DAMA can, on its own, precisely quantify the level of *TheHeap* background, just after a simple modification to the readout logic. In addition, we recommend comprehensive monitoring of helium concentration in the local environment, within the detector, and within the PMTs.

In the DM-Ice detector at the South Pole, the low operational temperature suppresses thermionic emission – one of the two components of *TheHeap* mechanism. Therefore, the coincidence rate of *TheHeap* events should be suppressed by at least four orders of magnitude compared to DAMA. DM-Ice has indeed reported an absence of seasonal variations. The other component of *TheHeap* mechanism is the afterpulsing rate per electron. Modern photosensor technology [3-5] can both suppress the afterpulsing rate and confine the afterpulses to a very narrow time interval: this would result in at least four orders of magnitude suppression of coincident *TheHeap* events with respect to PMTs, already at room temperature.

**ACKNOWLEDGMENT**

This material is based upon work supported by the U.S. Department of Energy, Office of Science, Office of High Energy Physics, under the Award DE-SC0018697 for the project *"The ABALONE™ Photosensor Panel Technology for Rare Decay and Rare Particle Detection."* Disclaimer: This report was prepared as an account of work sponsored by an agency of the United States Government. Neither the United States Government nor any agency thereof, nor any of their employees, makes any warranty, express or implied, or assumes any legal liability or responsibility for the accuracy, completeness, or usefulness of any information, apparatus, product, or process disclosed, or represents that its use would not infringe privately owned rights. Reference herein to any specific commercial product, process, or service by trade name, trademark, manufacturer, or otherwise does not necessarily constitute or imply its endorsement, recommendation, or favoring by the United States Government or any agency thereof. The views and opinions of authors expressed herein do not necessarily state or reflect those of the United States Government or any agency thereof.

# REFERENCES

1. R. Bernabei et al., DAMA/LIBRA Collaboration, First model independent results from DAMA/LIBRA-phase2, https://arxiv.org/abs/1805.10486 .
2. J.Va'vra, More on molecular excitations: Dark matter detection in ice, Phys. Lett. B 761(2016)58.
3. Daniel Ferenc, U.S. Patent 9,064,678, 2015. WO 2015/176000 A1.
4. Daniel Ferenc, Andrew Chang, Cameron Saylor, Sebastian Böser, Alfredo Davide Ferella, Lior Arazi, John R. Smith, Marija Šegedin Ferenc, ABALONE$^{TM}$ Photosensors for the IceCube Experiment, Nuclear Instruments And Methods In Physics Research A, in press; https://arxiv.org/abs/1810.00280 .
5. Daniel Ferenc, Andrew Chang, Marija Šegedin Ferenc, The Novel ABALONE Photosensor Technology: 4-Year Long Tests of Vacuum Integrity, Internal Pumping and Afterpulsing; https://arxiv.org/abs/1703.04546 .
6. V. Walia, F. Quattrocchi, H. S. Virk, T. F. Yang, L. Pizzinoband, B. S. Bajwa, Radon, helium and uranium survey in some thermal springs located in NW Himalayas, India: Mobilization by tectonic features or by geochemical barriers?, Journal of Environmental Monitoring 7(9)(2005)850-5.
7. Willy Dyck, B.Tanab, Seasonal variations of helium, radon, and uranium in lake waters near the Key Lake uranium deposit, Saskatchewan, Journal of Geochemical Exploration, 10 (1978) 153.
8. R. Bernabei et al., DAMA/LIBRA Collaboration, Performances of the new high quantum efficiency PMTs in DAMA/LIBRA, Journal of Instrumentation 7(2012)P03009.
9. M. B. Shah, D. S. Elliott, P. McCallion, and H. B. Gilbody, Single and double ionisation of helium by electron impact, J. Phys B21(1988)2751.
10. R. Mirzoyan, D. Müller, J. Hose, U. Menzel, D. Nakajima, M. Takahashi, M. Teshima, T. Toyama, T. Yamamoto, Evaluation of novel PMTs of worldwide best parameters for the CTA project, Nuclear Instruments And Methods In Physics Research A 845(2017)603.
11. E. Barbosa de Souza et al., DM-Ice Collaboration, First search for a dark matter annual modulation signal with NaI(Tl) in the Southern Hemisphere by DM-Ice17, Phys. Rev. D95(2017)032006. https://arxiv.org/abs/1602.05939 .